\documentclass[12pt]{iopart}

\usepackage{graphicx} 
\usepackage[spanish,english]{babel}
\usepackage[utf8]{inputenc}
\begin{document}

\selectlanguage{english}

 \title{An experiment to address conceptual difficulties in slipping and rolling problems}

\author{Alvaro Suárez$^1$, Daniel Baccino$^1$, Arturo C. Mart\'{\i}$^2$}

\address{$^1$Departamento de F\'{i}sica, Consejo de Formación en Educación, Uruguay}
\address{$^2$ Facultad de Ciencias, Universidad de la Rep\'{u}blica, Uruguay}

\ead{alsua@outlook.com} 
\ead{dbaccisi@gmail.com}
\ead{marti@fisica.edu.uy}

\date{\today}

\begin{abstract}
  A bicycle wheel that was initially spinning freely was placed in
  contact with a rough surface and a digital film was made of its
  motion. Using Tracker software for video analysis, we obtained the
  velocity vectors for several points on the wheel, in the frame of
  reference of the laboratory as well as in a relative frame of
  reference having as its origin the wheel`s center of mass. The
  velocity of the wheel`s point of contact with the floor was also
  determined obtaining then a complete picture of the
  kinematic state of the wheel in both frames of reference.
  An empirical approach of this sort to problems in mechanics can contribute to overcoming the considerable difficulties
  they entail.
\end{abstract}

\selectlanguage{english}

\maketitle 

\textbf{Introduction.} Problems involving rolling without slipping are
frequently included in mechanics courses at upper secondary and
introductory university  levels. Although the problems appear
to be simple, they pose a number of difficulties for students because
they require simultaneous application of static and dynamic
concepts. Studies on rolling without slipping in the field of\textit{
  Physics Education Research} confirm these difficulties. Some
studies demonstrate, for instance, that students have difficulty in
recognizing that the direction of the static frictional force on a
body that is rolling without slipping does not necessarily oppose the
direction of rolling \cite{carvalho2005rotation}. They also have
difficulty determining the direction of velocity at different points
on the wheel \cite{rimoldini2005student}.
 
Although such problems are frequently dealt with theoretically in
textbooks \cite{halliday2010fundamentals}, they are seldom studied
experimentally. The absence of an empirical approach may have a
negative effect on learning processes because students may be led to
think that the theoretical model has no connection with real life,
without actually having tested their preconceptions against
experimental evidence.  In terms of experimental approaches, 
it is worth mentioning the study of  the velocity of
the center of mass of a cylinder during rolling with and without
slipping by means of video analysis \cite{de2014video}
and proposal focused on the transition between slipping and rolling
\cite{suarez2019video}.

Here we describe an experiment to study the velocity distribution of the
wheel  viewed both from the laboratory frame of reference and from a
relative frame of reference with its origin at the wheel's center of
mass. The wheel initially is spinning freely, and is then put in contact
with a rough surface, so that eventually it is rolling without
slipping. This analysis enabled students to visualize and gain
intuition into the kinematics of a rigid body that is rolling with and
without slipping.

\begin{figure}[h]
\centering
\includegraphics[width=.350\textwidth]{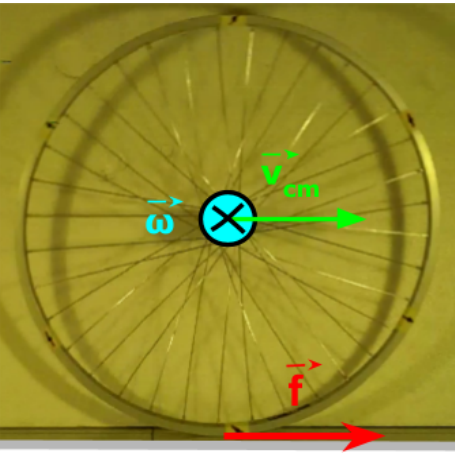}
\caption{A wheel is made to spin at an angular velocity $\omega_0$ and is then placed in contact with the floor. Once the wheel is released, the center of mass begins to accelerate due to the force of kinetic friction acting in the direction of movement.
}
\label{fig01}
\end{figure}

\textbf{Experimental setup.} The system consists of a bicycle wheel of
radius  as depicted in Fig.~\ref{fig01}. The
rim was marked at equal intervals in order to facilitate automatic
tracking of the wheel’s motion with Tracker software \cite{brown2012tracker}. Motion was
recorded at 30 \textit{fps} using a digital video camera (Kodak
PlaySport) mounted on a tripod in such a way that its optical axis was
at right angles to the plane of movement of the wheel. In order to
obtain the clearest possible image, we used spotlights to improve
luminosity and increase the camera's shutter speed.  We shall assume
that the mass of the wheel is symmetrically distributed around its
physical axis.

 The wheel, initially spinning is then placed on a horizontal floor and 
the velocity of center of mass, initially equal to zero, increases with time, due to the force of kinetic friction acting in the direction of movement.  This situation takes place over a very short time interval until  the velocity of the point of contact with the floor is zero in
the laboratory frame of reference. Then, the wheel begins to roll without
slipping so that both the velocity of the center of mass and the
angular velocity remain constant. As can readily be deduced from the equations of motion, when the wheel rolls
without slipping, the resultant external force must be zero and
therefore the static frictional force (the only force acting in the
direction of the horizontal axis) must also be zero. In practice the
wheel will not keep on rolling indefinitely because over longer
timespans, the rolling resistance that we here chose to neglect, must
be taken into account.

\begin{figure}[h]
\centering
\includegraphics[width=.450\textwidth]{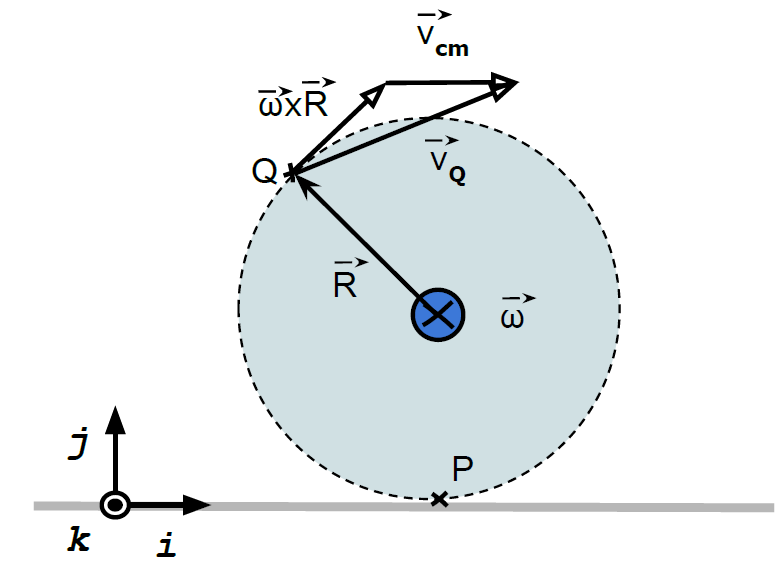}
\caption{Diagram of the velocity vector of a point $Q$ on the rim of
  the wheel represented as the addition of the vectors of rotation and
  translation. The point $P$ in contact with the floor is also
  indicated.  }
\label{fig02}
\end{figure}

\textbf{Velocity distributions.} The velocity at a given point on the
rim can be thought, as shown in Fig.~\ref{fig02}, as the addition of
two vectors, one translational, in which all the points on the wheel
have the same velocity as the center of mass, and one rotational,
where the direction of the velocity of each point on the rim is
tangential to the rim. 

In the laboratory frame of reference, while the wheel is slipping, the
velocity of the centre of mass is increasing and the angular velocity
is decreasing, the velocity of the point $P$ in contact with the floor
is opposite in direction to the wheel's motion and it decreases until
it becomes zero when the wheel is rolling without slipping. On the
other hand, when viewed in the frame of reference fixed to the wheel's
center of mass, point $P$ describes a circular motion that is
uniformly decelerated until the wheel is rolling without slipping,
when $P$ begins to describe uniform circular motion.  

\begin{figure}[h]
\centering
\includegraphics[width=.450\textwidth]{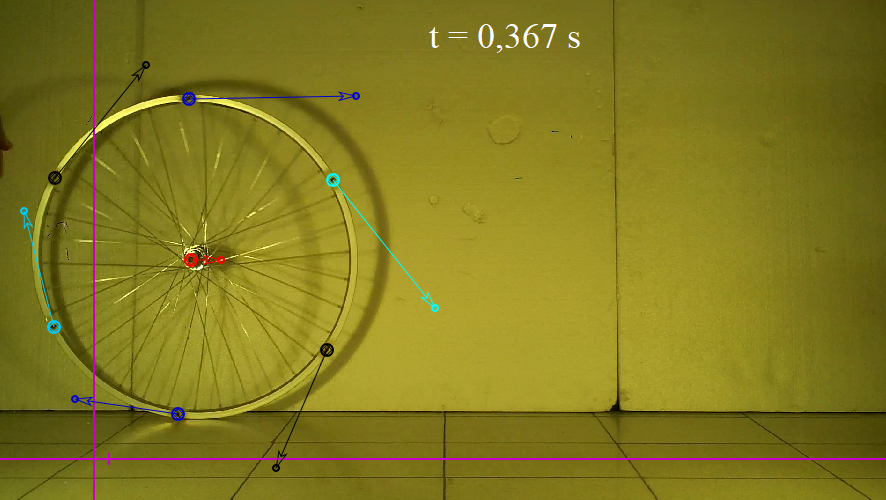}
\includegraphics[width=.450\textwidth]{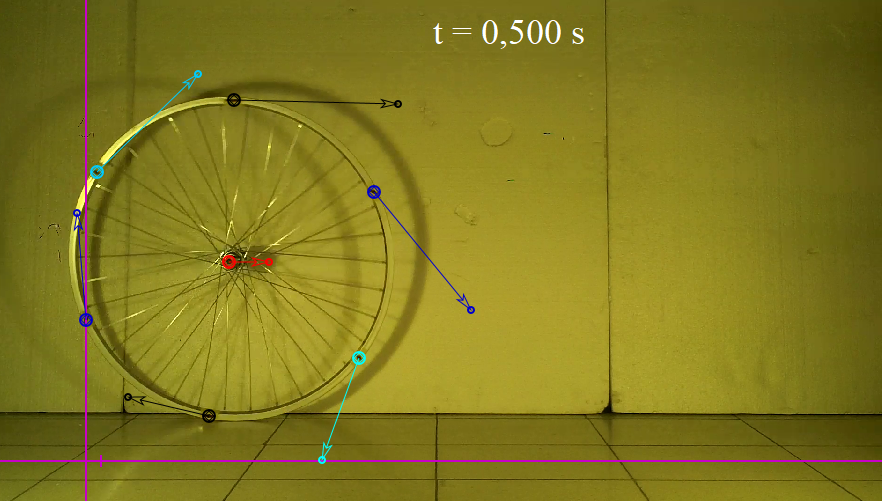}
\includegraphics[width=.450\textwidth]{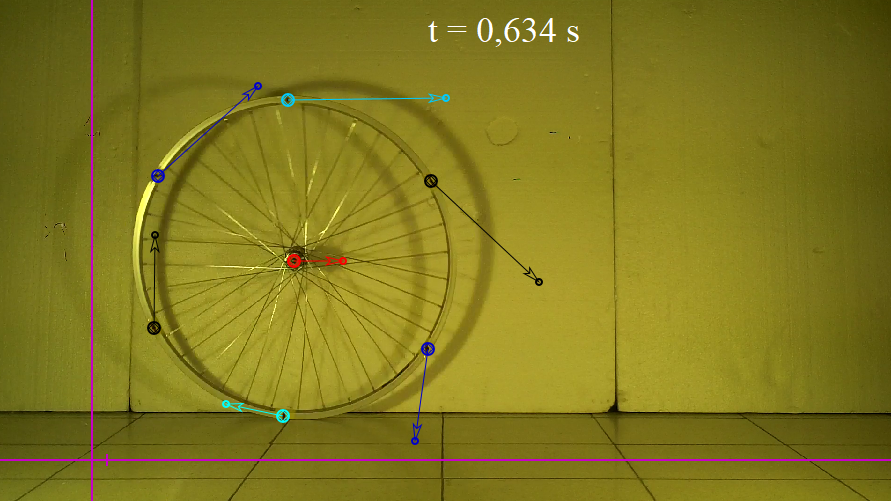}
\includegraphics[width=.450\textwidth]{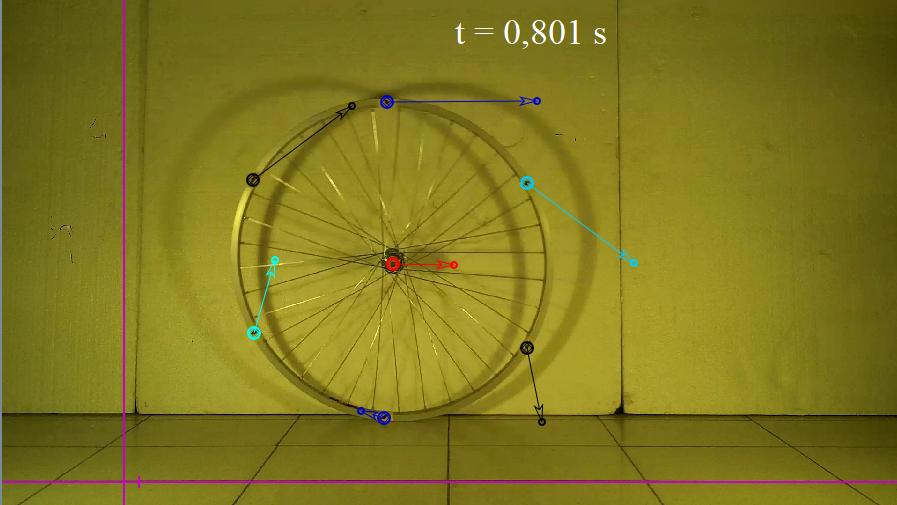}
\caption{Velocity vectors  corresponding to several points of the wheel
(drawn with different colours) in the laboratory frame of reference during slipping motion. Elapsed time is indicated in each panel.}
\label{fig04}
\end{figure}

\textbf{Experimental results} We used Tracker to analyse the changing
velocities of six points on the rim  from the time
slipping started until rolling without slipping was established. We
did this in the laboratory frame of reference and in a frame of
reference with its origin at the wheel`s center of mass.
Figure~\ref{fig04} shows a sequence of four images in which are
represented the velocity vectors at different points on the rim,
during the period when slipping is occurring and in the laboratory
frame of reference. As expected, the direction of the velocity of the
point of the wheel in contact with the floor is opposite to that of
the center of mass, and its modulus decreases as time elapses.

\begin{figure}[h]
\centering
\includegraphics[width=.40\textwidth]{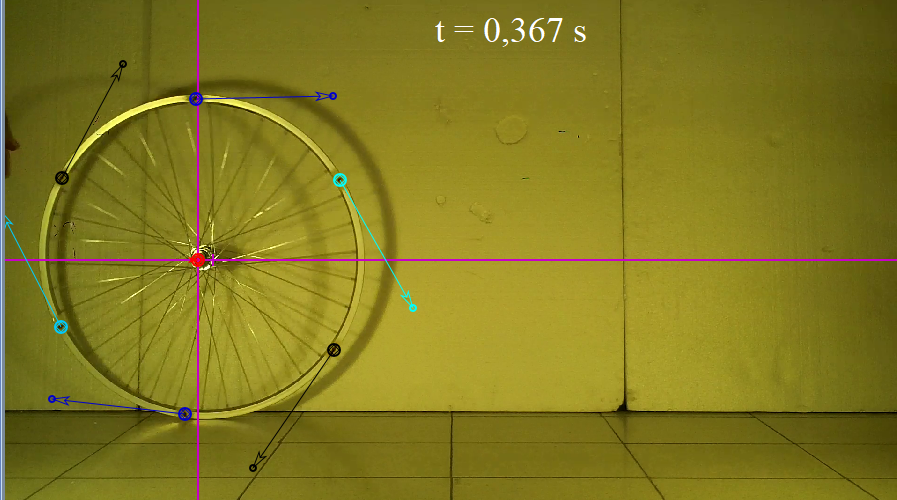}
\includegraphics[width=.40\textwidth]{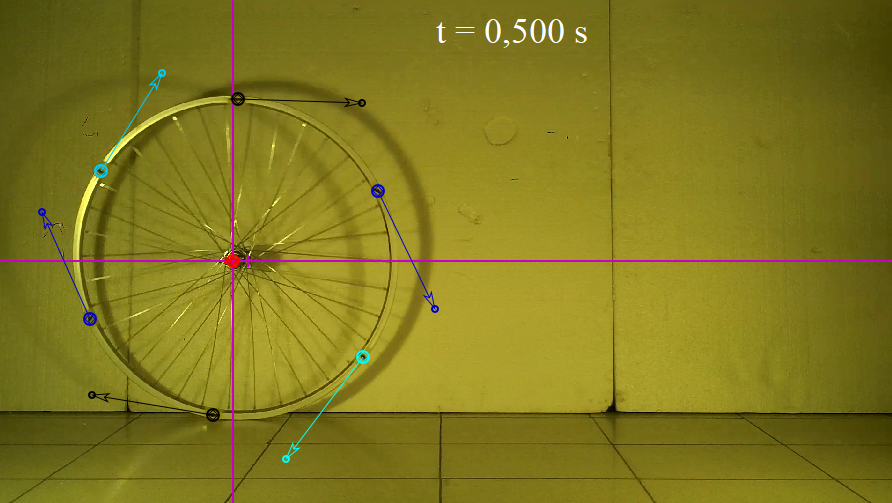}
\includegraphics[width=.40\textwidth]{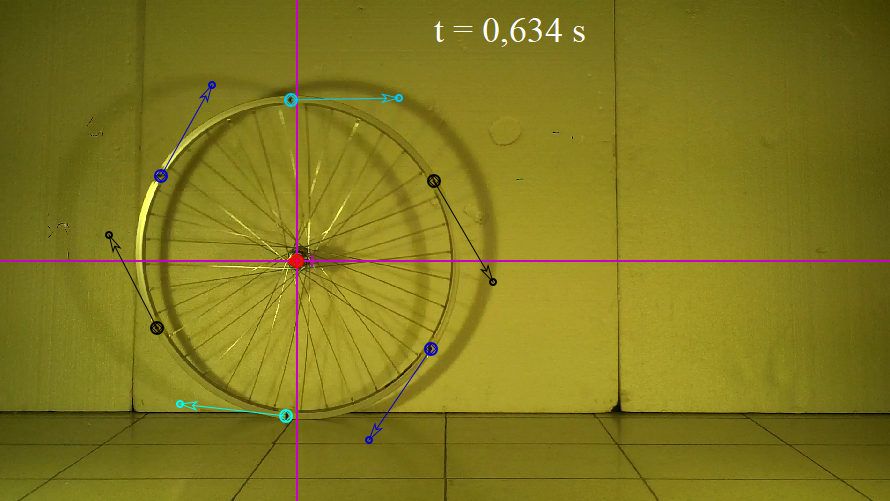}
\includegraphics[width=.40\textwidth]{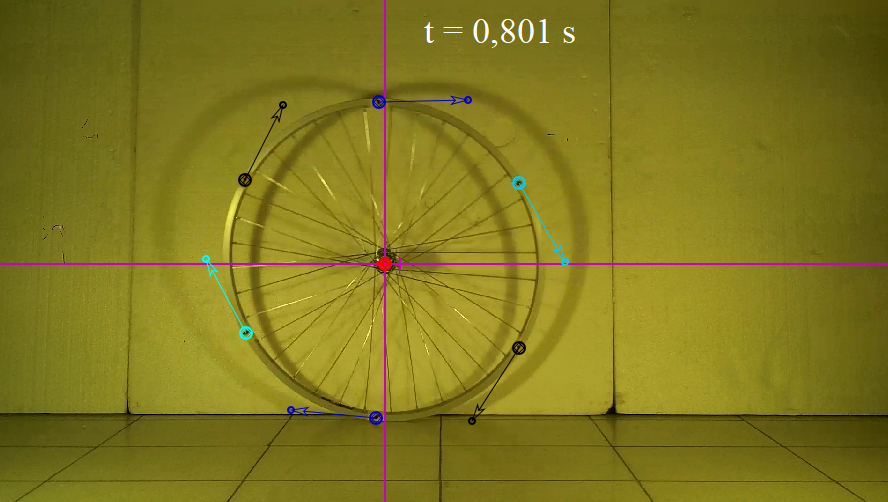}
\caption{Similar to Fig.~\ref{fig04} but the velocity vectors are
  measured with respect to the frame of reference fixed to the hub of
  the wheel.  }
\label{fig05}
\end{figure}

Making the wheel`s center of mass the origin of a coordinate system,
the sequence of images in Fig.~\ref{fig05} shows how the profile of
the velocities develops over time during the slipping phase in the
frame of reference centred on the hub. In contrast to the previous
figure, in all these images the velocity vectors are tangent to the
rim and in each snapshot are approximately equal in magnitude. It can
be also appreciated, in accordance with the theory, that the modulus
of the velocity diminishes as time elapses.

\begin{figure}[h]
\centering
\includegraphics[width=.440\textwidth]{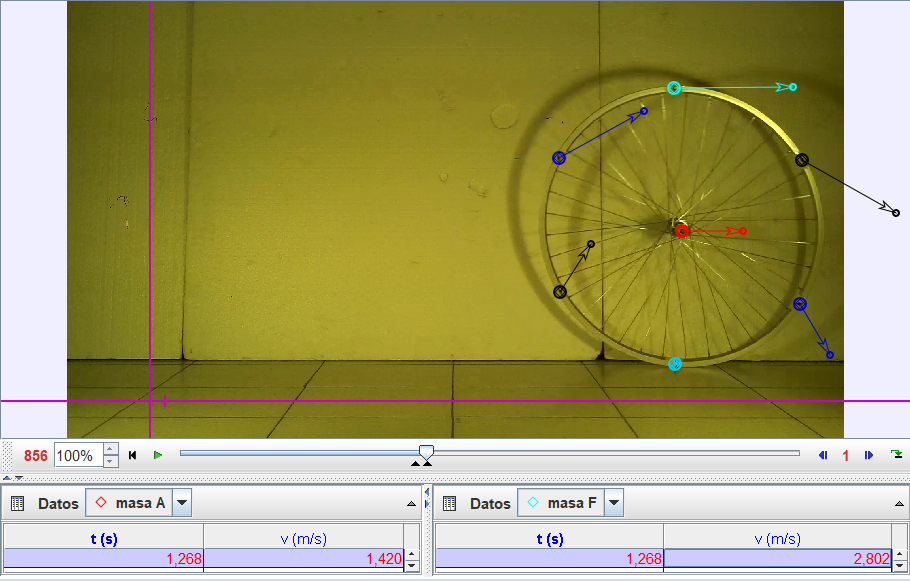}
\includegraphics[width=.440\textwidth]{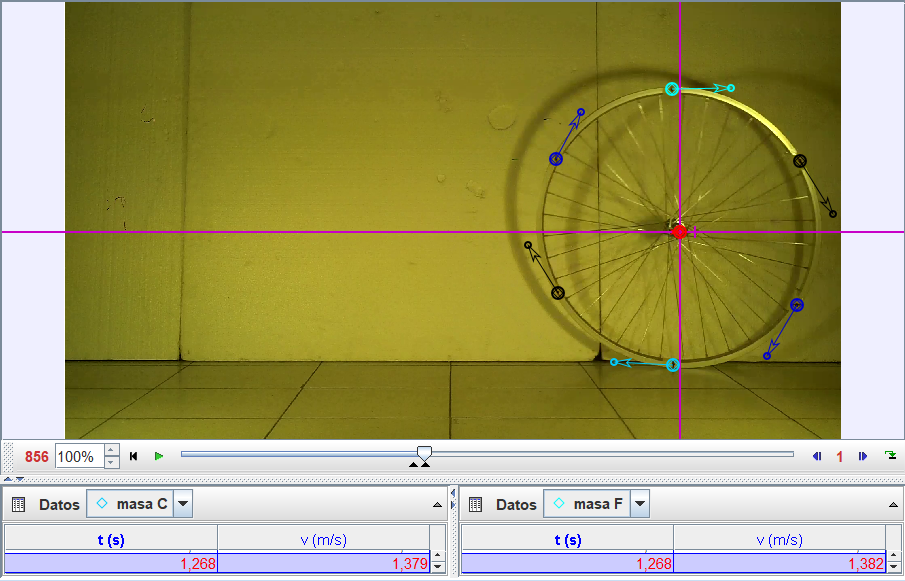}
\caption{Tracker screen snapshots showing in different colours the
  velocity vectors at different points on the wheel with respect to
  the laboratory frame of reference (left) and  with respect to the
  frame of reference with its origin at the center of mass (right).
  Below each panel, the values of the modulus of
  the velocity of the centre of mass ($A$) and of the highest point
  ($F$) of the wheel are indicated.  }
\label{fig06}
\end{figure}

To gain further insight, Fig.~\ref{fig06} shows the velocities
of the same points, in the laboratory frame of
reference (left) and in the frame of reference centred on the hub of the
wheel (right), but this time when the wheel is rolling without
slipping. Direct inspection of the left panel shows, as theory
predicted, that the velocity of the point of contact with the floor is
zero, while the velocity of the highest point is twice
that of the centre of mass \cite{halliday2010fundamentals}. On the
other hand, the right panel shows that the modulus of the velocity at
the highest and lowest points are similar to each other and similar,
also, to the modulus of the velocity of the centre of mass in the
laboratory frame of reference.

\textbf{Conclusion.} We studied the changes in velocities over time of
several points on the rim of the wheel with respect to the two frames
of reference described above. By analysing these changes, the students
are able to visualize and better comprehend what the directions of the
velocity vectors are at different points of the wheel. They are able
to observe, for example, how when the wheel is slipping the velocity
of the point of the wheel in contact with the floor is in the opposite
direction to motion, and they could see visual evidence for the
direction of action of the kinetic frictional force. At the same time,
they are able to comprehend the nature of the motion of the wheel from
the point of view of a frame of reference centered on the wheel hub
(the center of mass). This reinforced the idea of thinking about the
set of velocities of points in the laboratory frame of
reference as the addition of two velocity vectors, one translational
and one rotational. This concept is fundamental to the understanding
of rolling without slipping. Given all of the above, the analysis of
the changes over time of the set of velocities can be a very powerful
tool to help students overcome certain conceptual difficulties
associated with the kinematics of a wheel that is rotating with and
without slipping.

\section*{References}

\bibliographystyle{unsrt}
\bibliography{/home/arturo/Dropbox/bibtex/mybib}

\end{document}